# Demonstration of long-term thermally stable Silicon-Organic Hybrid Modulators at 85 °C

**Clemens Kieninger,**[1,2] **Yasar Kutuvantavida,**[1] **Hiroki Miura,**[3] **Juned N. Kemal,**[1] **Heiner Zwickel,**[1] **Feng Qiu,**[4] **Matthias Lauermann,**[1] **Wolfgang Freude,**[1] **Sebastian Randel,**[1] **Shiyoshi Yokoyama,**[3,4] **and Christian Koos**[1,2,*]

[1]*Karlsruhe Institute of Technology (KIT), Institute of Photonics and Quantum Electronics (IPQ), 76131 Karlsruhe, Germany*
[2]*Karlsruhe Institute of Technology (KIT), Institute of Microstructure Technology (IMT), 76344 Eggenstein-Leopoldshafen, Germany*
[3]*Department of Molecular and Material Sciences, Kyushu University, 6-1 Kasuga-koen Kasuga-city, Fukuoka 816-8580, Japan*
[4]*Institute for Materials Chemistry and Engineering, Kyushu University, 6-1 Kasuga-koen Kasuga-city, Fukuoka 816-8580, Japan*
[*]*christian.koos@kit.edu*

**Abstract:** We report on the first demonstration of long-term thermally stable silicon-organic hybrid (SOH) modulators in accordance with Telcordia standards of high-temperature storage. The devices rely on an organic electro-optic sidechain polymer with a high glass transition temperature of 172 °C. In our high-temperature storage experiments at 85 °C, we find that the electro-optic activity converges to a constant long-term stable level after an initial decay. If we consider a burn-in time of 300 h, the π-voltage of the modulators increases on average by less than 15 % if we store the devices for additional 2400 h. The performance of the devices is demonstrated by generating high-quality 40 Gbit/s OOK signals both after the burn-in period and after extended high-temperature storage.

**OCIS codes:** (250.7360) Waveguide modulators; (230.2090) Electro-optical devices; (130.3120) Integrated optics devices; (160.2100) Electro-optical materials; (200.4650) Optical interconnects;

## 1. Introduction

Silicon photonics is a promising integration platform for photonic devices, exploiting mature CMOS processes for low-cost high-yield mass fabrication of densely integrated photonic circuits that are ideally suited for high-volume applications such as optical communications [1]. However, due to the centro-symmetric crystal lattice of silicon (Si), the material does not exhibit a linear electro-optic (EO) effect [2]. For that reason, high-speed Si EO modulators are typically realized with reverse-biased pn junctions, exploiting the plasma dispersion effect [3]. Since the associated change in the refractive index is small, the π-voltage-length product $U_\pi L$ is typically as large as 10 Vmm [4–6]. This shortcoming of the Si platform can be overcome by combining silicon photonic or plasmonic waveguides with highly efficient organic EO (OEO) materials, leading to the so-called silicon-organic hybrid (SOH) [7,8] or plasmonic-organic hybrid (POH) [9,10] approach.

SOH devices have been demonstrated to show excellent performance. This includes, e.g., ultra-low voltage-length products down to $U_\pi L = 320$ Vµm [11] in combination with small efficiency-loss products down to $aU_\pi L = 1.2$ VdB [11], and high-speed modulation at line rates of up to 100 Gbit/s for on-off keying (OOK) [12], up to 120 Gbit/s for PAM4 [13], up to 160 Gbit/s for 40 GBd 16QAM [14], and 400 Gbit/s for 100 GBd 16QAM [15]. In addition, low chirp with an $\alpha$ - parameter of 0.1 [13] and a high extinction ratio in excess of 30 dB [16] have been shown. However, while these demonstrations outperform many competing device concepts in terms of efficiency, footprint, and speed, the reliability and long-term stability of SOH devices was less extensively investigated and still represents a weakness of the technology.

In this paper, we report on the first demonstration of long-term thermally stable SOH modulators that fulfill high-temperature storage requirements according to Telcordia standards GR-468-CORE [17]. The devices exploit a side-chain EO polymer with bulky adamantyl units that lead to a high glass transition temperature of $T_g = 172$ °C [18,19]. We investigate SOH devices stored at 85 °C in ambient atmosphere and show that the electro-optic activity reaches a constant long-term stable level after an initial decay. When allowing for a burn-in time of approximately 300 h, the average in-device EO activity decreases by less than 15 % for an additional 2400 h of high-temperature storage. At the end of the test period, the modulators exhibit a $U_\pi L$-product of 3.3 Vmm, which represents a four-fold improvement compared to previously demonstrated long-term stable OEO modulators [20,21]. To confirm that the high-temperature storage does not impair the high-speed performance of the modulator, we demonstrate the generation of high-quality 40 Gbit/s On-Off -Keying (OOK) signals both after the burn-in period and after extended high-temperature storage. We believe that our demonstration represents an important milestone towards industrial adoption of the SOH technology.

## 2. Silicon-organic hybrid device concept

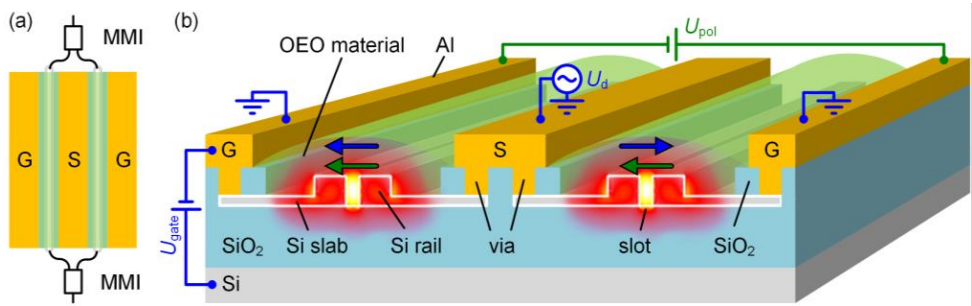

Fig. 1: Silicon-organic hybrid (SOH) device concept. **(a)** Top view of an SOH Mach-Zehnder modulator (MZM). The MZM features a ground-signal-ground (GSG) transmission line and two multi-mode interference (MMI) couplers for splitting and recombining the light in the two MZM arms. **(b)** Perspective view of a SOH MZM. Each MZM arm comprises a silicon (Si) slot waveguide consisting of two closely spaced Si rails that form a slot which is filled by the EO material. Electrical vias and thin *n*-doped Si slabs connect the GSG transmission line to the optical waveguide. Both the optical mode (red/yellow shading) and the radio frequency (RF) mode are highly localized in the slot region, which leads to efficient EO modulation. In order to establish macroscopic EO activity, the chip is heated to the glass transition temperature ($T_g$) of the OEO material, and a poling voltage $U_{pol}$ is applied across the ground electrodes. The corresponding poling fields (green arrows) induce an average acentric orientation of the dipolar EO molecules. This orientation is conserved by cooling the device and removing the poling field only when room temperature is reached. An applied signal voltage $U_d$ induces electric fields (blue arrows) which are parallel (antiparallel) to the poling direction in the left (right) arm of the MZM, enabling push-pull operation.

The concept of a silicon-organic hybrid (SOH) Mach-Zehnder modulator (MZM) is schematically depicted in Fig. 1, and Fig. 1(a) shows a top view of the device [13]. We use two multi-mode interference (MMI) couplers at the input and output of the MZM to split and recombine the light. The modulator electrodes form a coplanar transmission line in ground-signal-ground (GSG) configuration. Figure 1(b) shows a perspective view of the SOH MZM. Each MZM arm comprises an SOH phase shifter consisting of a Si slot waveguide. The slot waveguide is formed by two parallel 240 nm wide and 220 nm high Si rails separated by a 190 nm wide slot, which is filled by the OEO material. Electrical vias and thin *n*-doped Si slabs connect the aluminum (Al) GSG transmission line and the optical waveguides. An applied radio frequency (RF) voltage drops entirely across the highly resistive EO material in the slot. The optical mode, illustrated by a red and yellow shading in Fig.1 (b), is strongly confined to the slot region, resulting in a large overlap with the RF field in the OEO material. This leads to the excellent modulation efficiency of SOH devices. The silicon photonic chips are fabricated on standard 8-inch silicon-on-insulator (SOI) wafers in a commercial foundry using 248 nm optical lithography. The EO cladding is applied to the chips by an in-house post-processing step. To this end, the synthesized EO material is dissolved in the organic solvent cyclopentanone and deposited on the Si slot waveguides using a microdispenser. Subsequently, the coated chips are heated to remove remaining solvent from the EO material. Compared to spin coating, using the microdispenser allows for highly localized deposition of the material with a typical resolution of approximately 20 µm. In particular, the technique avoids covering the Al RF contact pads by the EO material and thus simplifies contacting the modulator with microwave probes. After deposition of the OEO material, the molecules are randomly oriented in the slot and a one-time poling process is required to establish average acentric molecular orientation leading to macroscopic EO activity. To this end, we heat up the chip close to $T_g$ of the OEO cladding material and apply a poling voltage $U_{pol}$ across the floating ground electrodes of the MZM. The voltage induces poling fields, indicated as green arrows in Fig. 1(b), in the slot regions of the MZM, which align the dipolar molecules in the OEO cladding. After cooling the device down to room temperature, the molecular orientation is frozen and the poling voltage can be removed. A drive voltage $U_d$ applied to the GSG transmission line induces electric fields in the two slot regions (blue arrows) that are parallel to the poling orientation in one arm and antiparallel to the poling orientation in the other arm of the MZM. This leads to phase shifts of equal magnitude but opposite sign in the two MZM arms, which results in an efficient and low-chirp push-pull operation of the modulator [13].

Regarding the high-speed performance of the devices, the main limitation is caused by the fact that the slot waveguide acts as a capacitor which has to be charged and discharged via the resistive doped Si slabs. The corresponding resistance-capacitance RC cut-off frequency can be increased by decreasing the Si slab resistance. This can be achieved by applying a gate voltage $U_{gate}$ between the Si substrate and the Si device layer which induces a highly conductive charge accumulation layer in the Si slabs [22]. For the high-speed experiments discussed in Section 4, we apply a gate field of about 0.07 V/nm, which results in a measured 3 dB EO bandwidth of 20 GHz. The bandwidth can be increased by using optimized doping profiles, which reduce the resistance of the Si slabs without compromising the low optical attenuation.

## 3. EO material and thermal stability tests of SOH modulators

For testing the stability and reliability of the SOH modulators, we adhere to Telcordia standards which specify generic test protocols for telecommunication equipment [17]. For modulators that are based on OEO materials, high-temperature storage tests over 2000 h at 85°C according Telcordia GR-468-CORE (section 3.3.2.1) are particularly challenging. This is due to the fact that the macroscopic EO activity of organic materials is externally induced in a poling process at elevated temperatures [23], which leads to an acentric orientation of the dipolar OEO molecules and hence to an increased potential energy of the system. This state is conserved by cooling the device to room temperature, thus reducing molecular mobility and preventing relaxation of the molecules towards an energetically favored random orientation. When exposing a poled OEO device to an elevated temperature, however, the molecular mobility is again increased, thus giving rise to a disordering of the molecules and to partial or complete loss of the electro-optic activity [20,21,24,25]. A critical parameter regarding thermal relaxation of acentric molecular order is the glass transition temperature $T_g$: Thermal relaxation is negligible as long as the storage temperature is well below $T_g$ and only becomes relevant for temperatures close to or above $T_g$. High-$T_g$ materials are hence instrumental for long-term thermal stability of modulators using OEO materials.

In our devices, we use a recently introduced side-chain EO material [18]. The molecular structure of the OEO material is depicted in Fig. 2(a). The material is based on a methyl methacrylate (MMA) polymer backbone with four different units. The first unit contains an adamantyl side group (magenta). Due to its bulky character, this side group makes the polymer chain less mobile and thus increases $T_g$. Additionally, the side group is nonpolar such that no detrimental interaction with the chromophore group arises. The second unit is a non-functionalized MMA group, and the third unit contains a phenyl vinylene thiophene (PVT) chromophore (blue) that is connected to the MMA backbone via a linker (red), and which leads to the EO activity of the polymer. The last unit is an unintentional residual, which contains only a linker side group where no PVT chromophore was attached. The sequence of the four units along a polymer chain is random, and the relative proportions amount to $l = 0.4$, $l = 0.4$, $m$ and $n$, for the first, second, third and fourth unit, respectively, where $m + n = 0.2$ and m >> n. For this composition, the material exhibits a high $T_g$ of 172 °C, measured by differential scanning calorimetry. The material thus promises excellent thermal stability even at elevated storage temperatures. During material synthesis, the proportion of all units can be adjusted by the concentration of the associated educts which allows to design both $T_g$ and the EO activity. For example, a very similar EO polymer with a higher adamantyl proportion of 54 % was recently synthesized and featured an increased $T_g$ of 194 °C but a slightly reduced EO activity [21].

To pole our specific EO polymer Fig. 2(a), its temperature is set to $T_g = 172$ °C while we apply an electric field of 140 V/µm. To quantify the modulation efficiency of the poled devices, we measure the π-voltages $U_\pi$ by using the experimental setup depicted in Fig. 2(b). An external-cavity laser provides the optical carrier that we couple to and out of the device via grating couplers (GC). Using a function generator (FG), we feed the modulator with a triangular electrical voltage having a peak-to-peak amplitude $U_{d,pp} > U_\pi$. The modulated out-coupled light is detected with a photodiode (PD), and an oscilloscope records both the modulated optical signal (red line, left axis) and the triangular drive voltage (blue line, right axis), see Fig. 2(c) for typical curves. Since the modulating voltage is chosen slightly larger than $U_\pi$, we can directly measure $U_\pi$ as the drive voltage difference, which switches the MZM between maximum and minimum transmission.

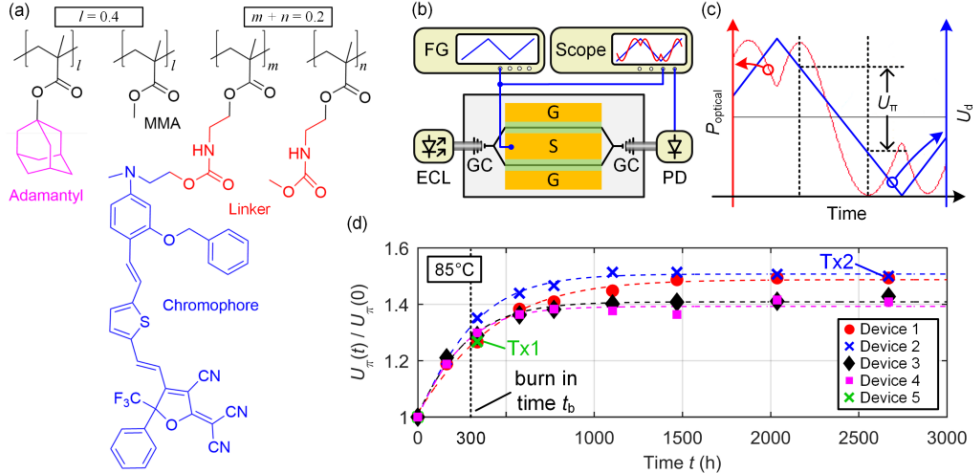

Fig. 2: EO material characterization. **(a)** Molecular structure of the EO polymer. The polymer chain consists of a random concatenation of many methyl-methacrylate-(MMA-)based units, marked by square brackets $[\text{unit}]_{l,m,n}$. The first unit contains a bulky *adamantyl* side group (magenta), which increases $T_g$ of the polymer, while the second unit is an MMA unit *without* a side group. The third unit contains a phenyl vinylene thiophene *chromophore* side group (blue), which induces the EO activity and which is connected to the MMA via a *linker* group (red). The fourth unit contains only the *linker* side group and corresponds to unintentionally unoccupied chromophore binding sites. The relative proportions of the four types of units (*adamantyl*, *without*, *chromophore*, *linker*) amount to $l = 0.4$, $l = 0.4$, $m$, $n$, respectively, with $m + n = 0.2$ and $m \gg n$. With this composition, $T_g$ amounts to 172 °C, measured by differential scanning calorimetry. **(b)** Experimental setup for static π-voltage measurement. Light of an external-cavity laser (ECL) is coupled to and out of the modulator by grating couplers (GC). A triangular waveform from a function generator (FG) is fed to the GSG transmission line. The modulated light is detected by a photo diode (PD). The signals of the PD and the FG are monitored by an oscilloscope. **(c)** Modulated optical power (red line, left axis) and triangular drive voltage (blue line, right axis) as a function of time. The device is operated at its 3 dB point. The voltage increment required for a transition from maximum to minimum transmission corresponds to the π-voltage $U_\pi$. **(d)** Long-term thermal stability testing of four SOH modulators at 85 °C according to Telcordia protocols GR-468-CORE (Section 3.3.2.1) for high temperature storage. We depict the normalized π-voltage as a function of time. In all experiments, the π-voltage reaches a constant long-term stable level after an initial increase. When allowing for a burn-in time of $t_b \approx 300$ h, the π-voltage of the devices increases on average by less than 15 % for an additional high-temperature storage period of at least 2400 h, i.e., $U_\pi(t_b + 2400 \text{ h}) < 1.15 \cdot U_\pi(t_b)$. Tx1 and Tx2 indicate devices that are used for data transmission experiments to prove that high-temperature storage does not impair the high-speed functionality of the devices, see Section 4. Device 1 … 4 were simultaneously tested in a first storage run, whereas Device 5 (Tx1), indicated by a green cross, was measured during a second run and removed from the oven shortly after the burn-in time $t_b$ to serve as a benchmark to Device 2 (Tx2) in the transmission experiments.

For a systematic investigation of thermally induced relaxation in SOH modulators, we pole four nominally identical devices and store them in an oven at a temperature of 85 °C in accordance with pertinent Telcordia standards GR-468-CORE (Section 3.3.2.1) for high temperature storage [17]. Note that the devices were not hermetically sealed and thus not protected from oxygen or humidity. During the storage, the devices are neither optically nor electrically connected. Since thermal relaxation reduces the in-device EO activity of the modulators, the π-voltage of the devices increases. To monitor this process, the modulators are removed from the oven from time to time to measure the change in $U_\pi$. The experimental results for the four devices are summarized in Fig. 2(d) where $U_\pi(t)$ normalized to its respective initial value $U_\pi(0)$ is plotted as a function of time $t$. All devices show qualitatively the same trend where after an initial increase during the first few hundred hours $U_\pi(t)$ converges towards a constant long-term stable level.

**Table 1: Summary of measurement and fitting results for thermal relaxation in SOH modulators**

| Device # | Experiment | | Fit | | |
|---|---|---|---|---|---|
| | $U_\pi(0\text{ h})$ [V] | $U_\pi(2700\text{ h})$ [V] | $a$ | $b$ | $\tau$ [h] |
| 1 | 1.54 | 2.30 | 0.672 | 0.313 | 329 |
| 2 | 1.48 | 2.22 | 0.662 | 0.336 | 236 |
| 3 | 1.52 | 2.17 | 0.710 | 0.281 | 217 |
| 4 | 1.54 | 2.17 | 0.718 | 0.281 | 199 |

For a better understanding of the relaxation processes, we model the decay of the EO activity by a theoretical model. There exists a multitude of published models to describe the thermally induced decrease in EO activity in poled organic EO materials. These models range from theory-based Debye models [26], which describe a simple exponential decay, to semi-empirical models describing a bi-exponential decay [27,28], or purely empirical models, e. g., stretched exponential models [29–31]. In this work, we choose a modified Debye model due to its simplicity and its good fit to the measured data. Taking into account that the π-voltage $U_\pi$ is inversely proportional to the EO activity and that it converges towards a stable value, the relaxation can be modelled by

$$\frac{U_\pi(t)}{U_\pi(0)} = \frac{1}{a + be^{-t/\tau}}. \qquad (1)$$

In this relation, the quantities $a$, $b$ and $\tau$ are used as fit parameters to adapt the model to the experimental findings. The quantity $1/a$ is the limit of $U_\pi(t)/U_\pi(0)$ for large times $t$, $b$ is a weighting parameter for the exponential decay of the EO activity, and $\tau$ is the corresponding characteristic decay time. The experimental data in Fig. 2(d) show good agreement with this model as indicated by the dashed lines, which are obtained from least squares fits. The increase of $U_\pi$ may be attributed to a thermally induced relaxation process in the EO polymer, which releases mechanical stress that was previously induced during the cooling step of the poling process. We believe that this effect can be further reduced by using optimized poling and cooling protocols. The fact that $U_\pi$ reaches a long-term stable level can be attributed to the large difference between $T_g = 172$ °C and the storing temperature of 85 °C, at which the molecular mobility of the polymer chain is negligible. Quantitatively, we deduce from Fig. 2(d) that after an initial burn-in time $t_b$ of approximately 300 h, the π-voltages of the devices increase on average by less than 15 % for at least additional 2400 h. In a potential application scenario, the poled SOH MZM would undergo the burn-in process after poling and prior to shipping the devices. It is therefore particularly interesting to compare the high-speed performance of a device right after the burn-in process with a device that has undergone the full 2700 h of high-temperature storage. To that end, we performed a second high-temperature storage run with a single device (Device 5) indicated by a green cross in Fig. 2(d), which we removed from the oven at $t = 330$ h, i.e., shortly after the burn-in time $t_b$. This device serves as a benchmark to Device 2 in the transmission experiments described in Section 4. The relative increase of $U_\pi$ of Device 5 is in good agreement with the other data sets, confirming the reproducibility of the experiments.

To the best of our knowledge, these experiments represent the first demonstration of SOH MZM that are long-term thermally stable in accordance with Telcordia standards for high-temperature storage. The measured values for $U_\pi(0\,h)$ and $U_\pi(2700\,h)$ as well as the fitting parameters for all four devices are summarized in Table 1 and indicate consistent performance. After 2700 h at 85 °C, the devices with a length of $L = 1.5$ mm exhibit an average π-voltage-length product $U_\pi L$ as low as 3.3 Vmm. This is roughly four times lower compared to other long-term stable organic-based EO modulators [20,21].

## 4. Data transmission experiment

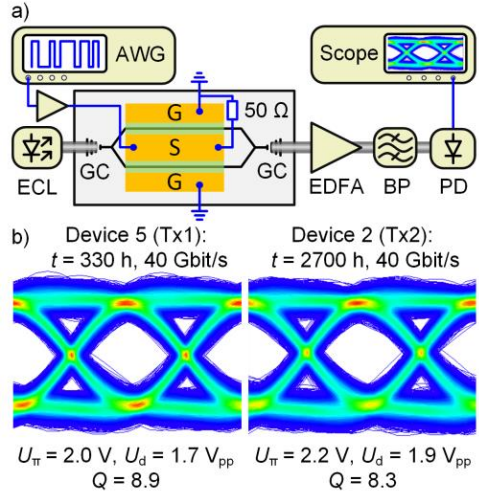

Fig. 3: Data transmission experiment. **(a)** Experimental setup. The electrical 40 Gbit/s signal is obtained from an arbitrary waveform generator (AWG). The drive signal is amplified and coupled to the chip by a microwave probe. A 50 Ω termination avoids back-reflection of the RF signal. An external-cavity laser (ECL) provides the optical carrier, which is coupled to and from the chip via grating couplers (GC). The modulated light is amplified in an erbium-doped fiber amplifier (EDFA), filtered by a bandpass filter (BP) and detected by a high-speed photodiode (PD) connected to a real-time oscilloscope. An adaptive digital filter at the receiver flattens the end-to-end frequency response of the system. **(b)** Measured 40 Gbit/s eye diagrams for Device 5 (left) after 330 h high-temperature storage and Device 2 (right) after 2700 h, see data points labeled as Tx1 and Tx2, respectively, in Fig. 2(d). For a similar signal quality, the drive voltage for Device 2 needs to be increased by approximately 10%. This increase corresponds to the 10% higher π-voltage of Device 2 at 2700 h as compared to Device 5 at 330 h.

To demonstrate the full functionality of our devices even after high-temperature storage, we validate the high-speed performance of the SOH MZM by generating 40 Gbit/s on-off keying (OOK) signals. To this end, we use a device shortly after its burn-in (Device 5), labeled by Tx1 in Fig. 2 (d), as well as a device that was stored for the full 2700 h (Device 2), labeled as Tx2 in Fig. 2(d). The experimental setup is shown in Fig. 3(a). We use an arbitrary waveform generator (AWG) as a signal source to obtain a pseudo random bit sequence of length $2^{15}-1$. The signal is amplified by a broadband RF amplifier and applied to the SOH MZM using a microwave probe. To avoid back-reflections of the RF signal, we terminate the modulator with an external 50 Ω impedance via another microwave probe. An external cavity laser (ECL) provides the optical carrier, which we couple to and from the chip via grating couplers (GC). The total insertion loss of the device is 14 dB and comprises fiber coupling losses, which amount to 4.5 dB per GC interface, and on-chip losses of 5 dB. The on-chip losses are caused by excess losses in passive structures such as strip waveguides, power splitters and strip-to-slot mode converters [32], which add up to about 2.5 dB, and losses in the phase shifter arms of the 1.5 mm long MZM, which amount to another 2.5 dB. The on-chip losses can be reduced on the one hand by an optimized design of the passive components and on the other hand, by lithography processes featuring a higher resolution that may result in lower sidewall roughness of the strip and slot waveguides. The high coupling losses can be significantly reduced by optically packaging the devices, e.g., by following the approach of photonic wire bonding [33]. Using this technique, indium phosphide lasers were connected to silicon photonic chips with coupling losses down to 0.4 dB [34]. The high insertion loss of the present devices requires compensation, and we amplify the optical signal after the chip using an erbium doped fiber amplifier (EDFA). To suppress amplified spontaneous emission noise we use a bandpass filter (BP). For signal detection, we use a high-speed photodiode connected to a real-time sampling oscilloscope. To flatten the end-to-end frequency response of the system, we apply an adaptive digital filter to the detected signal.

Figure 3(b) shows the recorded OOK eye diagrams at a data rate of 40 Gbit/s for Device 5 (left), which was tested after 330 h of high-temperature storage and Device 2 (right), which was tested at the end of the full 2700 h. After the respective storage times, the π-voltages amount to 2.0 V and 2.2 V for Device 5 and Device 2, respectively. The peak-to-peak drive voltages are chosen such that roughly the same signal quality is achieved for both modulators. For Device 5, the measured $Q$ factor amounts to 8.9 for a peak-to-peak drive voltage of 1.7 $V_{pp}$. For Device 2, we measure a $Q$ factor of 8.3 for a slightly higher peak-to-peak drive voltage of 1.9 $V_{pp}$. Thus, for achieving roughly the same signal quality for both modulators, we have to increase the drive voltage of Device 2 by about 10 % compared to the drive voltage of Device 5. This increase in drive voltage can be directly linked to the 10 % larger π-voltage of Device 2 after 2700 h high-temperature storage as compared to Device 5 after 330 h. These results indicate that high-temperature storage primarily results in the slightly increased π-voltage while high-speed performance remains unchanged.

## 5. Summary and outlook

We report on the first demonstration of long-term thermally stable silicon-organic hybrid (SOH) modulators in accordance with Telcordia standards of high-temperature storage. We find that after an initial burn-in time of 300 hours the SOH modulators retain more than 85 % of their modulation efficiency for at least an additional 2400 h. These demonstrations represent an important milestone towards industrial adoption of SOH technology outside a controlled laboratory environment. For future applications, the burn-in time can be significantly decreased by a burn-in at higher temperature, which accelerates the initial relaxation process [20]. The average π-voltage-length product of the four investigated 1.5 mm long modulators after 2700 h at 85 °C amounts to 3.3 Vmm. This corresponds to a four-fold improvement compared to previously reported long-term stable modulators based on organic EO materials [20,21]. Given the vast potential of theory-guided material optimization, we expect that even more efficient long-term stable SOH devices with $U_\pi L$ products below 1 Vmm will come into reach in the near future. We have also shown that the devices are suitable for applications in high-speed optical communication systems by demonstrating the generation of 40 Gbit/s OOK signals using devices shortly after the 300 h burn-in as well as devices that have undergone the full 2700 h of high-temperature storage at 85 °C. The results show that the long-term storage at 85 °C does not impair the general high-speed performance of the devices.

One of the last remaining challenges for an industrial adoption of SOH modulators is a potential photo-induced degradation of the organic EO chromophores due to the high optical intensities in the nanoscopic slot waveguide. Studies have shown that the degradation process is caused by an oxidation of the EO chromophores, and that it can be prevented by operating the modulators in an oxygen-free environment [20,35]. This can be achieved by, e.g., applying an oxygen-blocking encapsulation on the devices. We are thus confident that in the near future SOH modulators will be not only resistant against thermal relaxation but also against photo-induced damages.

## 6. Funding

Deutsche Forschungsgemeinschaft (DFG) project HIPES, 383043731; ERC Starting Grant 'EnTeraPIC', 280145; ERC Consolidator Grant 'TeraSHAPE', 773248; EU-FP7 project BigPipes, 619591; Alfried Krupp von Bohlen und Halbach-Stiftung; Helmholtz International Research School for Teratronics (HIRST); Karlsruhe School of Optics and Photonics (KSOP); Karlsruhe Nano-Micro Facility (KNMF); Cooperative Research Program of 'NJRC Mater. & Dev.' and 'Five-star Alliance'; Japan Society for the Promotion of Science (JSPS) KAKENHI Grant, JP266220712; Japan Science and Technology Agency (JST) CREST, 16815359.